\documentclass[prl,aps,twocolumn,superscriptaddress]{revtex4-2}
\usepackage{graphicx,color,setspace}
\usepackage[utf8]{inputenc}
\usepackage{mathtools}
\usepackage[normalem]{ulem}
\usepackage{tabularx}
\usepackage{enumerate}%
\usepackage{float}
\usepackage[normalem]{ulem}
\usepackage{comment}
\usepackage{soul}
\usepackage[caption=false]{subfig} %
\usepackage{ragged2e} %
\DeclareCaptionJustification{justified}{\justifying}

\usepackage{pifont}
\let\oldding\ding%
\renewcommand{\ding}[2][1]{\scalebox{#1}{\oldding{#2}}}%

\usepackage[labelfont=bf]{caption}

\begin{document}

\title{Inverse Thermodynamics: Designing Interactions for Targeted Phase Behavior}

\author{Camilla Beneduce}
\affiliation{Dipartimento di Fisica, Sapienza Universit\`{a} di Roma, P.le Aldo Moro 5, 00185 Rome, Italy}
\author{Giuseppe Mastriani}
\affiliation{Dipartimento di Fisica, Sapienza Universit\`{a} di Roma, P.le Aldo Moro 5, 00185 Rome, Italy}
\author{Petr \v{S}ulc}
\affiliation{School of Molecular Sciences and Center for Molecular Design and Biomimetics, The Biodesign Institute, Arizona State University, 1001 South McAllister Avenue, Tempe, Arizona 85281, USA}
\affiliation{School of Natural Sciences, Department of Bioscience, Technical University Munich, 85748 Garching, Germany}
\author{Francesco Sciortino}\
\affiliation{Dipartimento di Fisica, Sapienza Universit\`{a} di Roma, P.le Aldo Moro 5, 00185 Rome, Italy}
\author{John Russo}
\affiliation{Dipartimento di Fisica, Sapienza Universit\`{a} di Roma, P.le Aldo Moro 5, 00185 Rome, Italy}

\begin{abstract}
The traditional goal of \emph{inverse self-assembly} is to design interactions that drive particles toward a desired target structure. However, achieving successful self-assembly also requires tuning the thermodynamic conditions under which the structure is stable. In this work, we extend the inverse design paradigm to explicitly address this challenge by developing a framework for \emph{inverse thermodynamics}, i.e. the design of interaction potentials that realize specific thermodynamic behavior. As a step in this direction, using patchy particle mixtures as a model system, we demonstrate how precise control over both bonding topology and bond energetics enables the programming of targeted phase behavior. In particular, we establish design principles for azeotropic demixing and show how to create mixtures that exhibit azeotropy at any prescribed composition. Our predictions are validated through Gibbs-ensemble simulations [A.Z. Panagiotopoulos,
Molecular Physics {\bf 61}, 813-826 (1987)]. These results highlight the necessity of coupling structural design with thermodynamic engineering, and provide a blueprint for controlling complex phase behavior in multi-component systems.
\end{abstract}

\maketitle

\section{Introduction}

In atomic and molecular systems, interactions are fixed by quantum mechanics, limiting the ability to engineer specific thermodynamic behaviors. Soft matter, by contrast, thrives on tunability~\cite{preisler2016crystals,nguyen2017tuning,kumar2019inverse,dasgupta2020tuning,dwyer2023tunable,van2024soft}: interactions can be deliberately designed, whether through colloidal functionalization~\cite{sacanna2011shape,chakraborty2017colloidal,chintha2021modeling,chakraborty2022self,zhang2023shape}, DNA-mediated binding~\cite{diaz2020photo,mcmullen2021dna,liu2024inverse,zhou2024colloidal}, or anisotropic shapes~\cite{sacanna2011shape,kraft2011patchy,kraft2012surface,zhou2024space}. This flexibility has enabled the realization of exotic states of matter, such as equilibrium gels with vanishing coexistence regions~\cite{russo2009reversible,swinkels2024networks}, re-entrant phase transitions~\cite{sciortino2009phase,russo2011reentrant,roldan2013phase}, and empty liquids stabilized by limited valence~\cite{bianchi2006phase}. But beyond forward design (predicting properties from interactions), a deeper challenge emerges: the inverse thermodynamics problem. That is, how should interactions be tailored to achieve a desired thermodynamic outcome?

Significant strides have been made toward solving this problem~\cite{hagan2021equilibrium,dijkstra2021predictive,semwal2022tunable,kennedy2022self,marino2024crystallization,videbaek2024economical,shelke2024optimal,lu2025octo,de2025impact,frechette2025computer}. For instance, reducing the range of attractive potentials systematically depresses the critical temperature, eventually rendering the liquid-gas transition metastable~\cite{lekkerkerker1992phase,meijer1994colloids}. Similarly, finely tuned short-range repulsions can induce solid-solid transitions~\cite{hemmer2001solid}, while competing interactions generate re-entrant phenomena like melting upon cooling or condensation upon dilution~\cite{roldan2013gelling,rogers2015programming,bomboi2016re, hvozd2022empty}. These examples reveal general principles, but they often focus on qualitative trends (e.g., shifting phase boundaries) rather than quantitative control (e.g., prescribing exact coexistence conditions).

Certain thermodynamic conditions are pivotal for functionality. In equilibrium gelation, for example, the narrowing of the coexistence region (a hallmark of ``empty liquids''~\cite{russo2022physics}) suppresses phase separation, enabling gelation without interference of a glass transition~\cite{sciortino2017equilibrium,russo2022physics,swinkels2024networks}. For self-assembly, azeotropy, a condition where liquid and vapor phases share identical composition, plays an analogous role. Azeotropes act as thermodynamic attractors during nucleation: by fixing the composition of coexisting phases, they maximize the yield of target structures when the azeotropic point matches the stoichiometry of those structures. In our prior work~\cite{beneduce2023two,beneduce2023engineering}, we demonstrated that azeotropicity accelerates self-assembly by matching liquid and crystal composition. Here, we focus on the inverse problem: how can interactions be designed to place an azeotropic point at any desired concentration?

To achieve this, we focus on binary mixtures of patchy particles~\cite{russo2024synthesis}, a paradigm for programmable matter. Patchy particles are coarse-grained models for systems having specific, short-ranged, and directional interactions such as DNA functionalized colloids~\cite{seeman2017dna,xiong2020three,li2025cocrystals} and DNA-origami~\cite{zhang20183d, park2019design,hayakawa2022geometrically,liu2024inverse,kahn2025arbitrary,saha2025modular}.
Patchy colloids offer unparalleled control over both bonding topology (via patch number and geometry) and interaction energetics (via patch strength and specificity~\cite{andreas2025}). Previous studies exploited this to assemble complex crystals ~\cite{pinto2024automating,pinto2024inverse,wang2025inverse}, but thermodynamic precision, such as engineering phase diagrams with prescribed azeotropes, remains underexplored.

Our results are derived within the framework of Wertheim’s perturbation theory, which provides a closed-form description of phase equilibria in associating fluids. These predictions are rigorously validated via Gibbs ensemble simulations, a method ideally suited for mapping phase diagrams of complex fluids with minimal computational cost. Unlike brute-force free energy calculations, the Gibbs ensemble directly samples coexistence conditions, bypassing the need for thermodynamic integration, a critical advantage for multi-component ~\cite{vega2008determination,de2011phase}.

We demonstrate how tuning patch interactions and bond energies can  position azeotropic points at targeted locations. As a striking example, we design mixtures that exhibit azeotropic demixing across all compositions, where dew and bubble curves (properly defined in the following) collapse into a single line, a condition we term an \emph{ideal azeotropy}.

\section{Methods}

We use patchy particles as the computational model.
Patchy particles are spherical hard core colloids of diameter $\sigma$ decorated, on their surface, by attractive sites named patches that have range $\delta/2$ and angular width $2\theta_{max}$. We describe their pair interaction through the Kern-Frenkel potential~\cite{kern2003fluid,bol1982monte}

\begin{equation}
	\label{eqn:KF}
	V(\mathbf r_{ij},\hat{\mathbf r}_{\alpha,i},\hat{\mathbf r}_{\beta,j})= V_{SW}(r_{ij})F(\mathbf r	_{ij},\hat{\mathbf r}_{\alpha,i},\hat{\mathbf r}_{\beta,j})
\end{equation}

\noindent
where $V_{SW}$ is a square-well of depth $\epsilon$ and width $\delta$. $F$ depends on the orientation of the particles

\begin{equation}
	\label{eqn:f_KF}
	F(\mathbf r_{ij},\hat{\mathbf r}_{\alpha,i},\hat{\mathbf r}_{\beta,j})=
	\begin{cases}
		1 &\text{if}\quad 
		\begin{array}{l}
			\hat{\mathbf r}_{ij} \cdot \hat{\mathbf r}_{\alpha,i} > \cos{(\theta_{max})}\\ 
			\hat{\mathbf r}_{ji} \cdot \hat{\mathbf r}_{\beta,j} > \cos{(\theta_{max})}
		\end{array} \\
		0 &\text{otherwise}
	\end{cases}
\end{equation}

\noindent where ${\mathbf r}_{i,j}$ is the center-to-center distance between particle $i$ and particle $j$ and $\hat{\mathbf r}_{\alpha,i}$ ($\hat{\mathbf r}_{\beta,j}$) indicates the position of patch $\alpha$ ($\beta$) of particle $i$ ($j$). 
Different species of patchy particles can differ in the number (valence), placement, type, range, and angular width of the patches. In the following, we consider cases where the species differ only in their patch type, sharing the same radius, too. Moreover, we note that the energy gain in forming a bond $\epsilon_{\alpha \gamma}$ generally depends on the patches $\alpha$ and $\gamma$ involved.

We perform Monte Carlo simulations with aggregation-volume-bias moves (AVB)~\cite{chen2000novel,rovigatti2018simulate} both in bulk conditions (NVT and NPT simualations) and in the Gibbs ensemble to study phase coexistence of AB binary mixtures.
To directly access chemical potentials in simulations we use the S0 method~\cite{cheng2022computing}, where the following Kirkwood-buff integral is evaluated

\begin{equation}\label{eqn:s0_method}
        \mu_A^{ex}(x_A) = k_B T \int_{1}^{x_A} \frac{d x_A}{x_A} \left[\frac{1}{S_{AA}^0 - S_{AB}^0 \sqrt{x_A/x_B}} \right]
\end{equation}

\noindent and then $\mu_A$ is obtained according to equation below

\begin{equation}\label{eqn:s0_method_all}
        \mu_A(x_A) = \mu_A^0 + k_B T \ln{\bigg( \frac{x_A}{x_A^0} \bigg)}  + \mu_A^{ex}(x_A)
\end{equation}

\noindent where $x_A$ ($x_B$) denotes the molar fraction of species A (B) and $\mu_A^{0}$ is the chemical potential at a standard molar fraction. We selected this reference to be the pure state $x_A=1$. Simulations in the NPT ensemble are first carried out for different molar fractions to compute the structure factors $S_{AA}$, $S_{BB}$, and $S_{AB}$. Then, the values of $S_{AA}^0$, $S_{BB}^0$, and $S_{AB}^0$ in $\mathbf{k} = 0$ are extrapolated by fitting the Ornstein-Zernicke relation~\cite{barrat2003basic}, given by:

\begin{equation}
    S_{AB}(\mathbf{k}) = \frac{S_{AB}^0}{1 + k^2 \xi_A \xi_B}
    \label{eqn:S}
\end{equation}

\noindent where $\xi_i$ is the correlation length of species $i$. Finally, by numerically evaluating the integral in Eq. (3), the excess chemical potential is calculated.

From a theoretical point of view, we describe patchy particle mixtures with Wertheim first-order perturbation theory~\cite{wertheim1984fluids,chapman1988phase,de2011phase,jonas2022extended}. Considering a reference system of hard spheres, the Helmholtz free energy per particle in units of $k_BT$ for a $n$ component mixture of patchy particles can be expressed as the sum of a reference and a bonding contribution:

\begin{equation}
\label{eqn:betaf_reference}
	\begin{array}{l}
                \beta f = \beta f_\text{reference} + \beta f_\text{bonding} \\[8pt]
			\text{where} \quad \beta  f_\text{reference}=\beta f_\text{ideal}+\beta f_\text{HS} \quad \text{with} \\[8pt]
			\quad \quad \quad \ \ \beta f_\text{ideal}=\ln{\rho}-1+ \sum\limits_{i=1}^n x^{(i)}\ln{(x^{(i)}V_i)} \\[8pt]
			\quad \quad \quad \ \ \beta f_\text{HS}=\frac{4\phi-3\phi^2}{(1-\phi)^2} \text{,} \quad \text{and} \\[8pt]
            \beta f_{bonding}=\sum\limits_{i=1}^n x^{(i)} \Biggl[\sum\limits_{\alpha\in\Gamma(i)}\bigg(\ln{X_{\alpha}^{(i)}}-\frac{X_{\alpha}^{(i)}}{2}\bigg)+\frac{1}{2}n(\Gamma(i))\Biggr] \\
            \quad
		\end{array} \\
\end{equation}

\noindent where $\rho$ is the density, $x^{(i)}$ is the molar fraction of species i, $V_i$ is the cube of the de Broglie thermal length, $\phi$ is the packing fraction equal to $\rho V_s$ where $V_s=\sigma^3 \pi / 6$ is the volume of a single particle, and $X_{\alpha}^{(i)}$ is defined by the law of mass action in Eq. (6) that we use to inverse design an azeotrope. We note that azeotropic solutions $X_\alpha^{(i)}=X$ formally reduce the free energy of the mixture to that of a single component system.

To draw binodal curves from Wertheim's theory we use the isochoric thermodynamics framework~\cite{deiters2017differential,bell2018construction,beneduce2023engineering}.

\section{Results}

\begin{figure}[!t]
    \centering
    \includegraphics[width=0.45\textwidth]{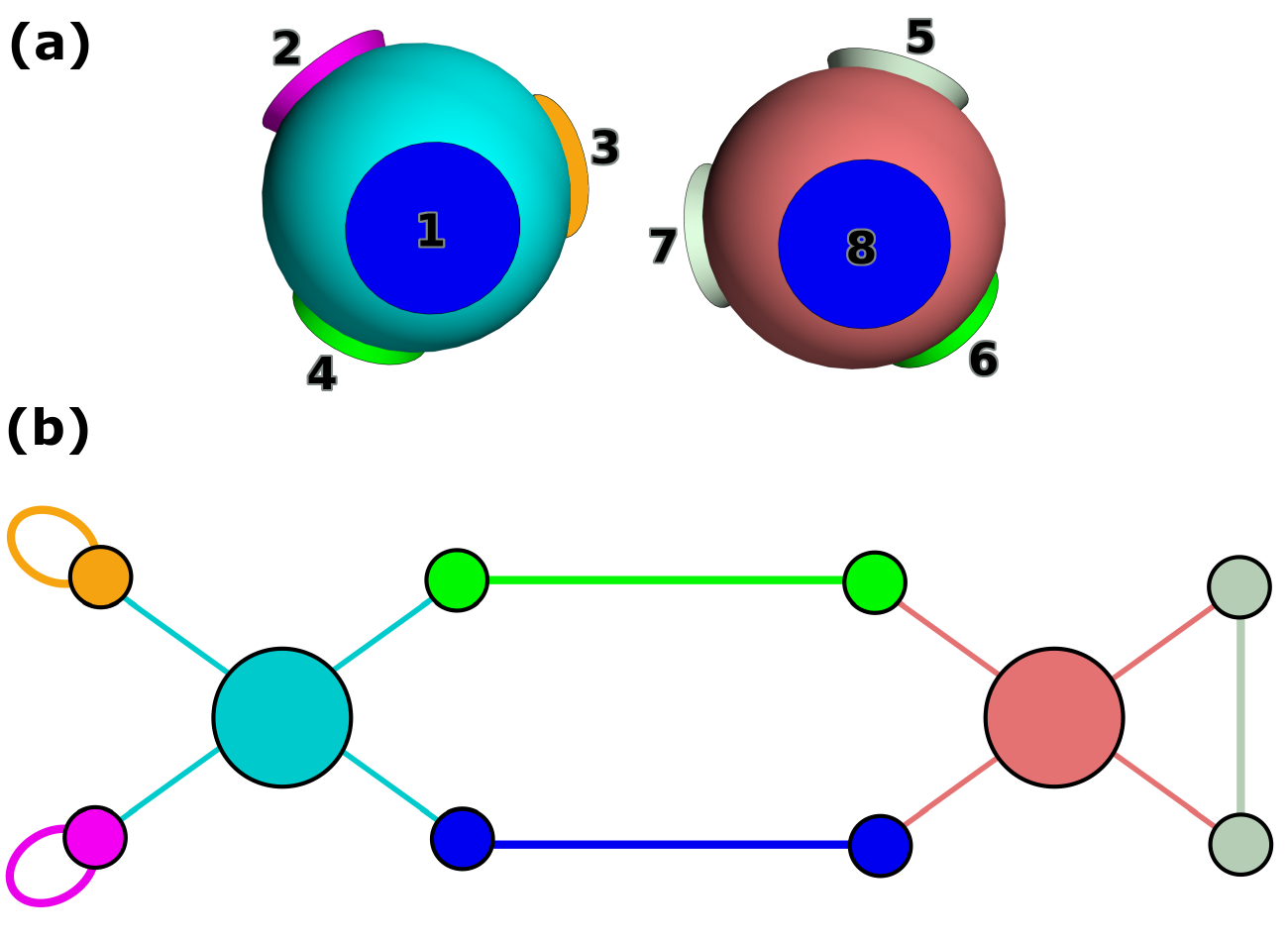}
    \caption{\textbf{N2c8 design.} a) 3D representation of the N2c8 design. The N2c8 system is a binary mixture of patchy particles, each featuring four patches tetrahedrally arranged. The first species, whose type of patches are labeled $1-4$, is colored cyan, while the second species, with type of patches labeled $5-8$, is colored red. The notation N2 refers to the two species, and c8 represents the eight distinct types of patches. Each pair of patches capable of forming bonds is represented by the same color. b) Graph of the N2c8 design. The two central nodes represent the two species, with each species connected to other nodes corresponding to its patches. Patches of same color indicate interacting patches. All the links are colored accordingly: links from each species to its patches are colored to match the species, while links between pair of interacting patches are colored to correspond with the patches.
    }
    \label{fig:graph_N2c8}
\end{figure}

Given a mixture of patchy particles, the design of the mixture is determined by the ways in which the patches can interact with each other and how they are distributed among the different components (patchy particle types).
The system's design is represented as a graph where nodes correspond to species, with each species node connected to its respective patch nodes, and where edges between patches indicate complementary binding interactions.
Self-loops represent self-complementary patches, i.e. patches that can bind to identical patches on other particles of the same species.
The weight of each edge measures the interaction energy of the bond. A graph is unweighted if all interactions have the same energy, or weighted otherwise.
As an example, in Fig.~\ref{fig:graph_N2c8} we plot the graph for the N2c8 mixture~\cite{rovigatti2022simple}: it is a binary mixture (N2 denotes two species) of patchy particles, each with four distinct patches tetrahedrally arranged (c8 refers to eight different patches).

In this framework, the inverse thermodynamics problem reduces to determining how to modify the graph topology, either through its edge configuration or edge weights, to achieve target thermodynamic properties. Here we specifically address azeotropic conditions, requiring the mixture to exhibit an azeotropic line (extending from P=0, T=0 to a binary critical point) at a predetermined concentration. This target concentration could correspond, for instance, to the stoichiometric ratio of the crystalline phase where nucleation rate is maximized~\cite{beneduce2023two}.

In the following we present both strategies that tune the bond connectivity (the list of edges) or the interaction energies  (the edge weights) to achieve control over the location of the azeotropic point.
Additionally, we analyze in detail the \emph{ideal azeotrope} condition, a special case where the mixture exhibits azeotropic demixing across all concentrations.

\subsection{Azeotropy from bond connectivity}\label{sec:connectivity}

The effects of bond connectivity on the thermodynamics can be quantified via the mass-action equation

\begin{equation}
\label{eqn:X}
X_{\alpha}^{(i)}=\biggl[ 1+ \phi  \sum_{j=1,N_s} x^{(j)} \sum_{\gamma\in\Gamma(j)} X_{\gamma}^{(j)} \Delta_{\alpha\gamma} \biggr]^{-1}
\end{equation}

\noindent $X_{\alpha}^{(i)}$ is the probability that patch $\alpha$ of species $i$ is not bonded, $N_s$ is the total number of species, $\Gamma(j)$ is the number of patches on species $j$, $\phi$ is the total packing fraction, $x_j$ is the molar fraction of species $j$, and $\Delta_{\alpha \gamma}$, being proportional to $e^{\epsilon_{\alpha \gamma}}$, accounts for the bond strength between patch $\alpha$ and $\gamma$.

At the azeotrope, the mixture behaves as a single-component system, and we can identify the condition where all patches $\alpha$ of each species $i$ have the same probability of being bonded (and equivalently of not being bonded) as a sufficient condition for the appearance of an azeotrope. In the Wertheim formalism this means that the mass balance Eq.~\ref{eqn:X} must be the same for all patches in the system: $X_{\alpha}^{(i)}= X$.

If all interaction energies are the same, $\Delta_{\alpha\gamma}\equiv\Delta$, and if we require that each patch has a unique bonding partner among all patches of all species in the mixture (the \emph{bond-exclusivity} condition~\cite{beneduce2023engineering}), then Eq.~\ref{eqn:X} becomes

$$
X_{\alpha}^{(i)}+\phi x^{(j)} [X_{\gamma}^{(i)}]^{2} \Delta + \phi (x^{(j)}-x^{(i)}) X_{\alpha}^{(i)} \Delta -1=0
$$

This equation admits the azeotropic solution $X_{\alpha}^{(i)}=X$ if $x^{(i)}=1/N_s$, \textit{i.e.} at equimolar conditions:

$$
\label{eqn:X_exclu_3}
X+\frac{\phi}{N_s} X^2 \Delta -1=0
$$

With the bond-exclusivity condition 
each patch has a unique and distinct bondable patch.
In a previous work~\cite{beneduce2023engineering}, we explicitly verified the presence of the azeotropic point at equimolar conditions by computing the pressure-concentration and density-concentration phase diagrams.

Azeotropy can be moved away from equimolar concentrations by allowing for multiple-bonding between patches~\cite{beneduce2023engineering}. In the following we focus on a special condition, called \emph{ideal azeotropy}, for which azeotropy occurs at all concentrations.

\subsection{Ideal Azeotropy}

The law of mass action (Eq.~\ref{eqn:X}) admits an azeotropic solution $X_{\alpha}^{(i)}=X$, that is independent of molar fractions when each patch has $N_s$ complementary patches, one for each of the $N_s$ species of the mixture.
In this case the equation becomes

\begin{equation}
\label{eqn:X_aa}
X+\phi X^2 \Delta -1=0
\end{equation}

\noindent that is independent of molar fractions $x_i$, i.e. the system has an azeotrope at every concentration $x_A\in[0,1]$.

\begin{figure}[!t]
    \centering    \includegraphics[width=0.35\textwidth]{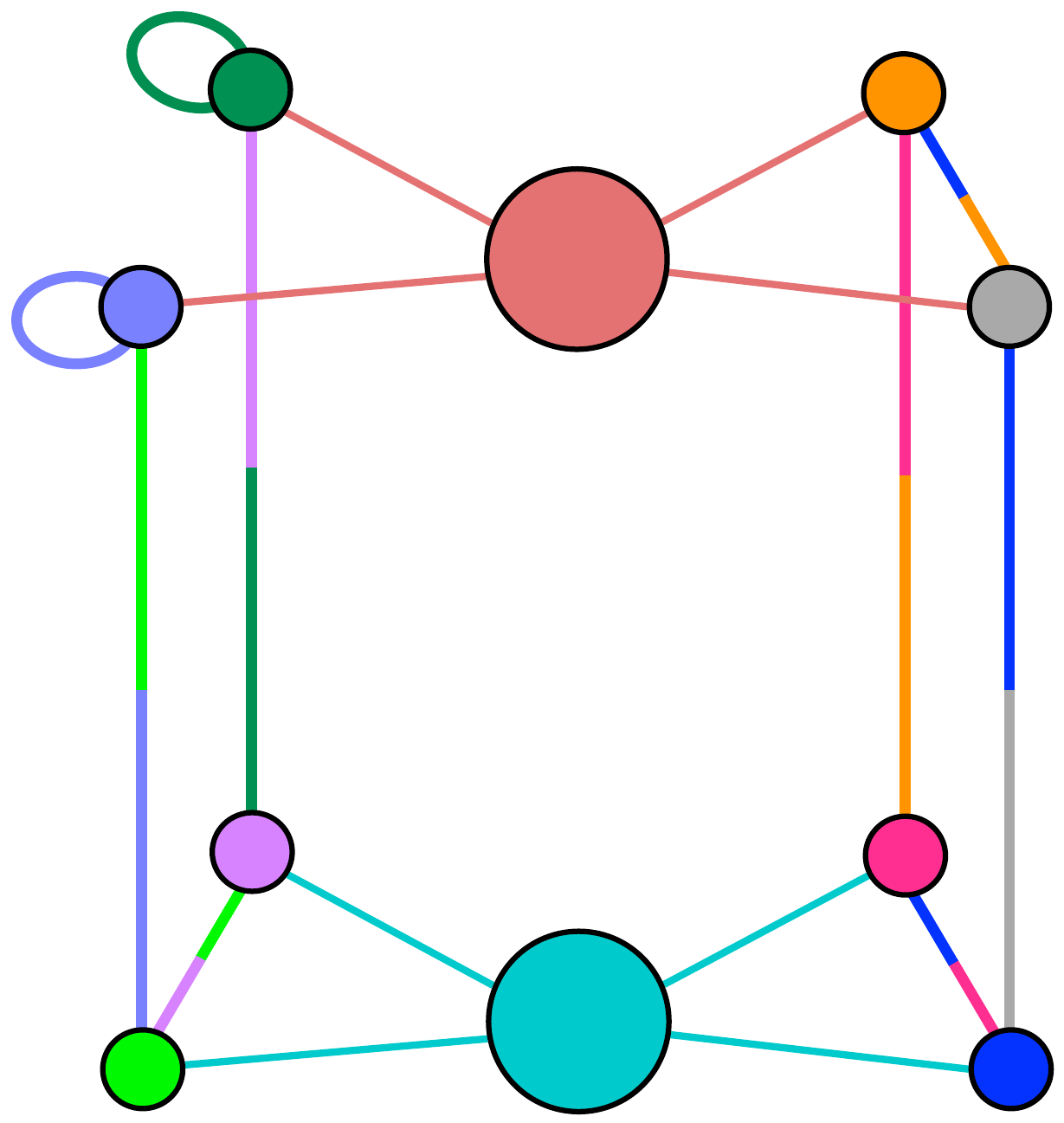}
    \caption{\textbf{Ideal azeotropic binary mixture.} The graph features two central nodes, representing the two species, colored cyan and red. Each node is connected by links of the same color to its respective patches. To emphasize that all patches are distinct, each vertex is uniquely colored. Bi-colored links between patches signify that each patch can connect to two other patches, one from the first species and the other from the second species.  Each vertex is incident to three links: two colored according to the patches it interacts with and the third one (cyan or red) indicating the species it belongs to.}
    \label{fig:ideal_azeotrope_graph}
\end{figure}

\begin{figure}[!t]
    \centering
    \includegraphics[width=0.52\textwidth]{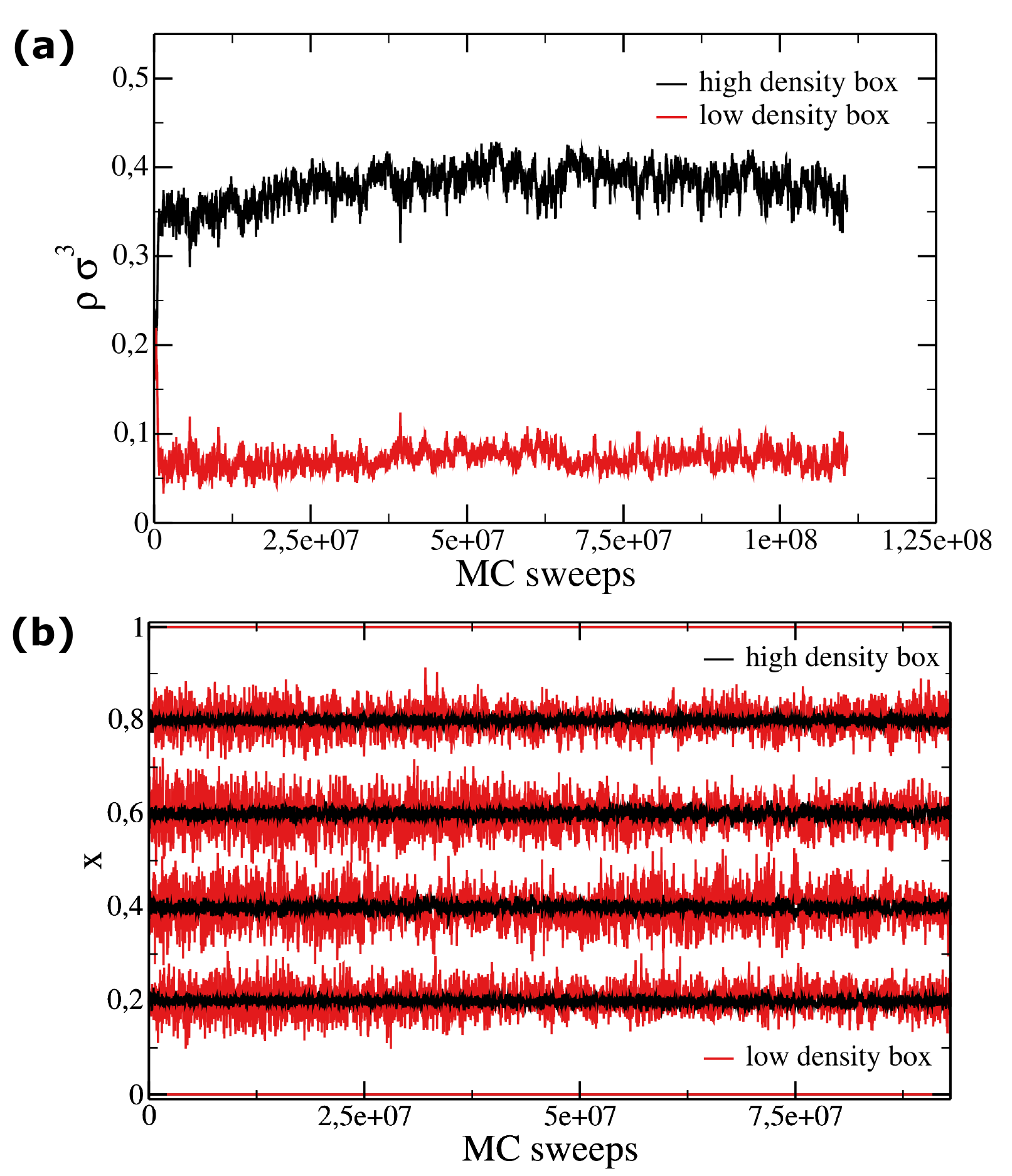}
    \caption{\textbf{Gibbs ensemble simulations of the always azeotropic mixture.} a) The two starting equal boxes with density $\rho=0.2$, N=$500$ particles, and temperature $T=0.138$ rapidly reach an equilibrium configuration with one box (liquid phase) denser than the other one (vapor phases). The figure shows just the $x=0.8$ case for clearness. b) Unlike density, the concentration values in both boxes remains constant throughout the simulation. The concentration of the vapor phase (depicted in red) exhibits greater fluctuations than that of the liquid phase (depicted in black) due to the smaller number of particles in the vapor box, but in both cases the initial concentration is preserved. This holds true regardless of the starting concentration, proving that the mixture is always azeotropic.}
    \label{fig:gibbs_density}
\end{figure}

A bond topology for a binary mixture that satisfies the ideal azeotropy requirements is shown in Fig.~\ref{fig:ideal_azeotrope_graph},
where each patch has two interacting patches, one on each specie.
To study de-mixing behavior of the design we perform Gibbs ensemble simulations~\cite{panagiotopoulos1987direct,panagiotopoulos1988phase}. We initialize two boxes at density $\rho=0.2$ by randomly placing $500$ particles in each. We explore the concentration range $x\in[0,1]$ with a step size of $\Delta x =0.2$ (where $x$ denotes the concentration of the first species A), and we run simulations over a temperature range from $T=0.125$ to $T=0.138$. Within this temperature range, coexistence between the two phases is observed. The density profiles for the case $T=0.138$ and $x=0.5$ are shown in Fig.~\ref{fig:gibbs_density}a, as an example: the two initially identical boxes equilibrate to different values, each corresponding to the density of the liquid and vapor phases.
Moreover, as shown in Fig.~\ref{fig:gibbs_density}b for the case $T=0.138$ and $x=0.8$, each concentration remains equal to its initial value in both boxes, resulting in a density-concentration phase diagram characterized by all tie-lines parallel to the density axis.

\begin{figure}[!t]
    \includegraphics[width=0.5\textwidth]{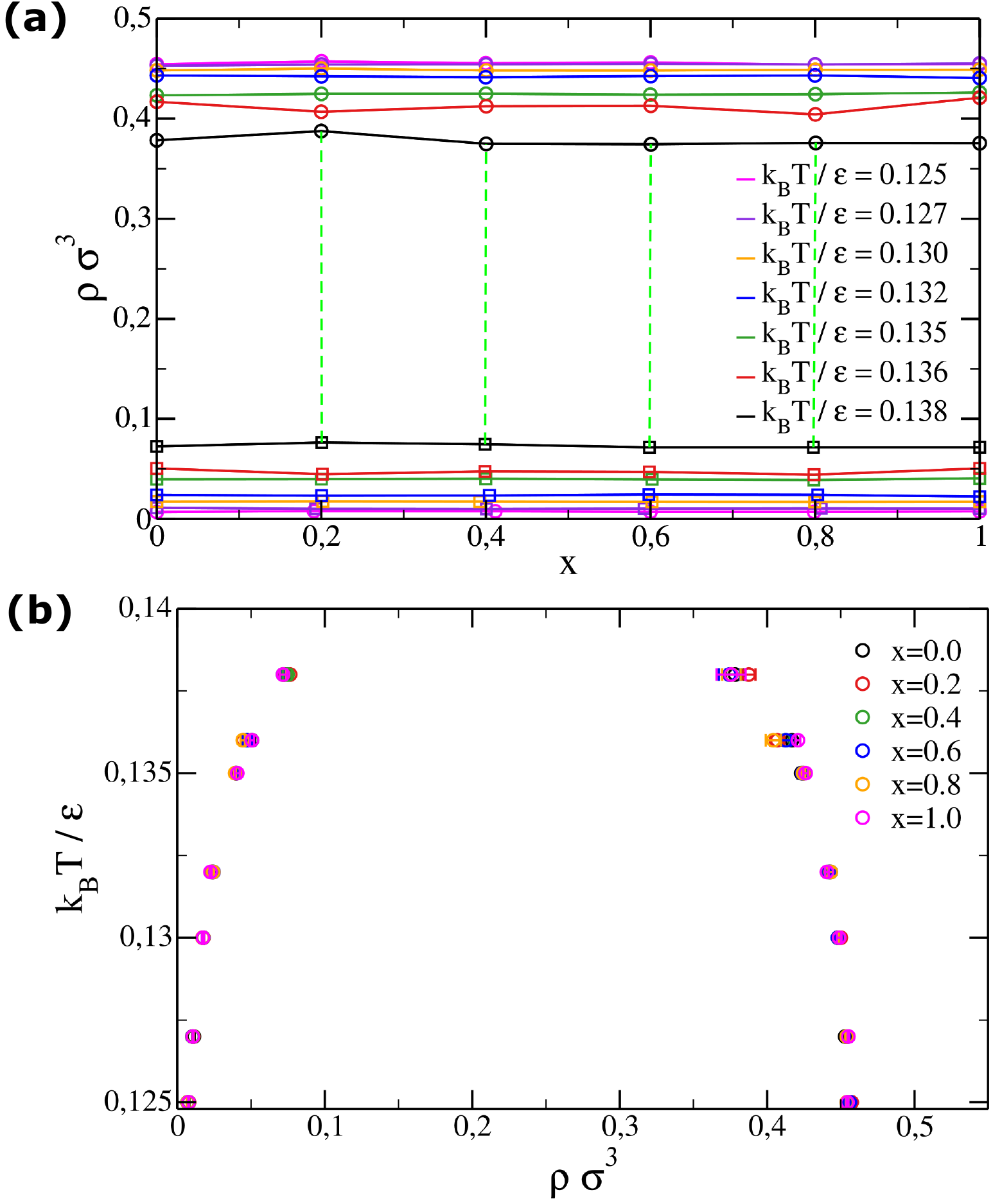}
    \caption{
    \textbf{Density concentration and temperature density phase diagrams of the always azeotropic binary mixture.} Monte Carlo simulations with AVB moves in the Gibbs ensemble are performed for each concentration $x$ of the first species, ranging from $0$ to $1$ in steps of $0.2$. The starting configuration consists of two equal boxes with density $\rho=0.2$ and N=$500$ particles. We study in parallel temperatures ranging from $T=0.125$ to $T=0.138$. Once equilibration is achieved and one box results in a liquid phase while the other in a gas phase, equilibrium density and concentration values are computed. (a) Density concentration phase diagrams. Although error bars are omitted for clarity, both the concentration and density errors are less than $10^{-3}$. The dew and the bubble point curve are parallel to the concentration axis and the tie lines (green dashed lines) are straight. Regardless of the starting concentration, the two coexisting phases maintain the same concentration value: the mixture is always azeotropic. (b)
    Temperature density phase diagrams. The coexistence region is independent from the concentration, confirming that the designed mixture always behaves as a single component system.}
    \label{fig:gibbs_phase_diagram}
\end{figure}

For each simulated temperature $T$ and each initial concentration $x$, we compute the coexistence densities and concentrations by averaging the values assumed by each box at equilibrium. The equilibrium densities become independent of concentration. Consequently, the density-concentration phase diagram, shown in Fig.~\ref{fig:gibbs_phase_diagram}a, is characterized by a bubble point curves (the locus of points where  vapour begins to coexist with the liquid phase as pressure is lowered starting from a point greater than the total
vapour pressure) and a dew point curve (the locus of points where liquid begins to coexist with the vapour phase as pressure is increased starting from a point in the vapour phase) both parallel to each other and to the concentration axis. 
The system's behavior as a single component system is evident also from the temperature concentration phase diagrams in Fig.~\ref{fig:gibbs_phase_diagram}b, which is plotted for each initial concentration of the two boxes. The overlapping curves  demonstrate the absence of concentration dependence.

\begin{figure*}[!t]
   \includegraphics[width=0.99\textwidth]{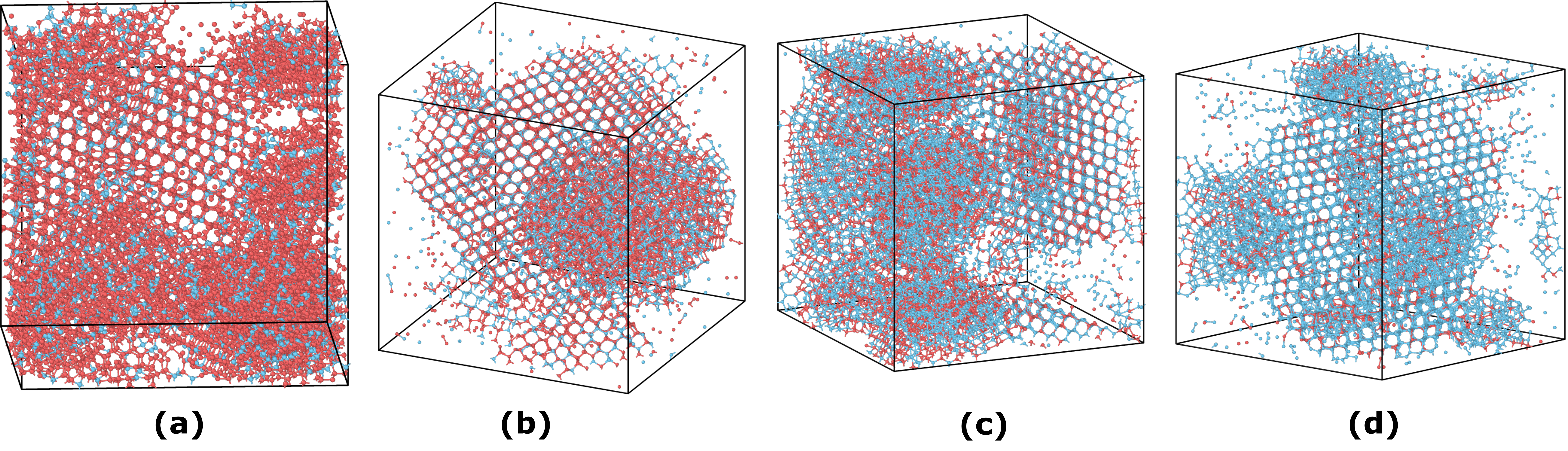}
    \caption{\textbf{Nucleation of the always azeotropic binary mixture.} Snapshots of successfully nucleated trajectories at different concentrations $x$ of the first species: $x=0.2$ in a), $x=0.4$ in b), $x=0.6$ in c), $x=0.8$ in d). Regardless of concentration, large crystal formation is observed, as each lattice position can be occupied equally by either a particle of the first species (blue colored) or of the second species (red colored). Monte Carlo simulations are performed with AVB moves in the canonical ensemble at temperature $T=0.109$, density $\rho=0.3$, and number of particles $N=1000$.}
    \label{fig:nucleation}
\end{figure*}

The concentration independence properties of ideal azeotropes also reflects in their crystallization behavior. To check this we run NVT simulations at $T=0.109$ and $\rho=0.3$. As illustrated by nucleation snapshots in Fig.~\ref{fig:nucleation}, crystallization, which occurs in the denser liquid phase after phase separation, involves substitutional solids where each lattice site occupied by one species can be replaced by the other. This explains why extended crystals can form at any concentration. These observations support the conclusion that the two species have equivalent bonding probabilities in the mean-field approximation and that the mixture behaves effectively as a single-component system.

\begin{figure}[!t]
    \centering
    \includegraphics[width=0.45\textwidth]{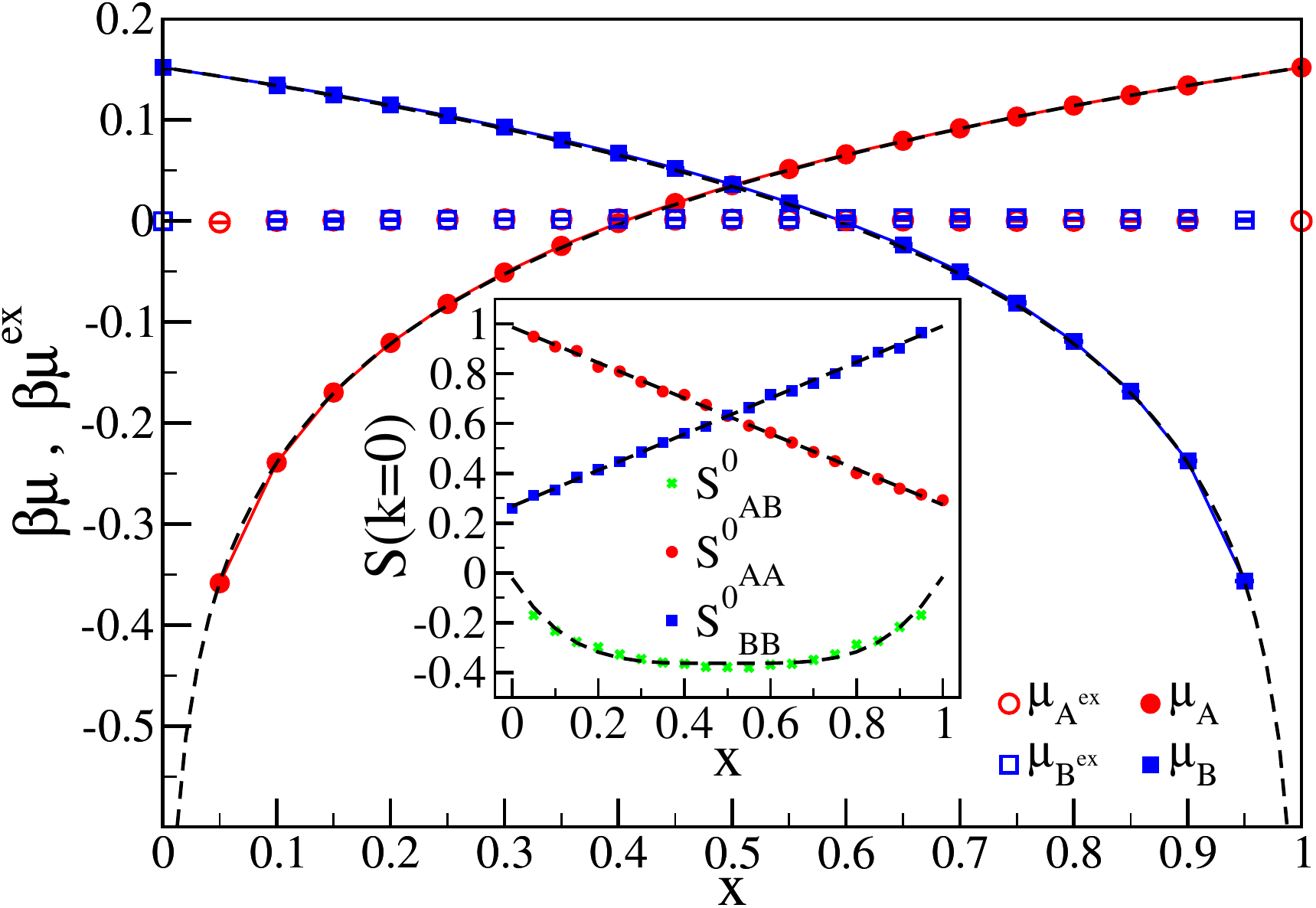}
    \caption{\textbf{S0 method results.} (Inset) The zero-wavenumber ({\bf k}=0) limit of the three partial structure factors $S_{AA}^0$, $S_{BB}^0$ and $S_{AB}^0$. The dashed lines indicate the analytic expression  fitted to the data and later on used to solve Eq.~\ref{eqn:s0_method} for $\mu^{ex}$. (Main) Total chemical potential $\mu$ (full symbols) and excess chemical potential $\mu^{ex}$ (empty symbols) for both species  highlighting the near-ideal behavior of the always azeotropic mixture.}
    \label{fig:s0_method}
\end{figure}

To further confirm the ideal character of the solution, we measure the chemical potential at different concentrations with the S0 methods~\cite{cheng2022computing} (also described in the Methods section) at one specific temperature. Simulations were performed in the NPT ensemble for species concentrations in the range $ x \in[0.05, 1]$. The simulation parameters used were: $N = 4000$ particles, pressure $P = 0.125$, and temperature $T = 0.17$. The system density for these parameters was found to be $\rho \sim 0.352$. The $\mathbf{k} = 0$ values of the determined structure factors $S_{AA}(\mathbf{k})$, $S_{BB}(\mathbf{k})$, and $S_{AB}(\mathbf{k})$ are shown in the inset of Fig.~\ref{fig:s0_method}, where the black dashed line represents the fit performed on the obtained data according to Eq.~\ref{eqn:S}.

The excess chemical potential was computed using Eq.~\ref{eqn:s0_method}, with uncertainties estimated from variations in the fitted partial structure factors shown in the inset of Fig.~\ref{fig:s0_method}.
To determine the standard chemical potential for pure components, we performed grand canonical ensemble simulations at $T=0.17$, systematically varying the chemical potential until the measured density matched that obtained from NPT simulations. This yielded a standard chemical potential of $\mu^{0} = 0.152$.
As shown in the main panel of Fig.~\ref{fig:s0_method}, the total chemical potential closely follows the ideal value (dashed line). This near-ideal behavior confirms the system's ideal azeotropic character, where composition changes minimally affect the excess chemical potential that constantly remains close to zero across all concentrations.

\subsection{Azeotropy from bond-energy control}

While in the previous sections we have demonstrated precise control over azeotropic composition through bond topology modifications, an alternative approach becomes essential when the bonding topology must remain constrained. In such cases, we can instead exploit the tunability of interaction energies to engineer the desired azeotropic behavior.

Two distinct phases (denoted as phase $'$ and phase $''$) coexist at a certain temperature and pressure when there is chemical equilibrium, \textit{i.e.}:

$$
    \mu_i' = \mu_i''
$$

In multicomponent systems, the chemical potential $\mu_i$ of each component depends on the system's composition. Since the mole fractions sum to unity ($x_1 + \cdots + x_n = 1$), only $n-1$ composition variables are independent.

For binary mixtures, we select one independent concentration variable (denoted $x$). The equilibrium condition requires:

\begin{align}
    d(\mu_i' - \mu_i'') &= 0 \nonumber\\
    \frac{\partial \mu_i'}{\partial x'} dx' - \frac{\partial \mu_i''}{\partial x''} dx'' &= 0 \nonumber
\end{align}

Along the azeotropic line, the compositions coincide ($x' = x''$ and $dx' = dx''$), simplifying the condition to:

$$
    \left.\frac{\partial \mu_i'}{\partial x}\right|_{x'} = \left.\frac{\partial \mu_i''}{\partial x}\right|_{x''}
$$

This equality of chemical potential derivatives, in addition to the equality of chemical potentials themselves, defines the special thermodynamic state of an azeotrope. Our algorithm exploits this relationship to control azeotropic composition.

\begin{figure}[!t]
    \centering    \includegraphics[width=0.45\textwidth]{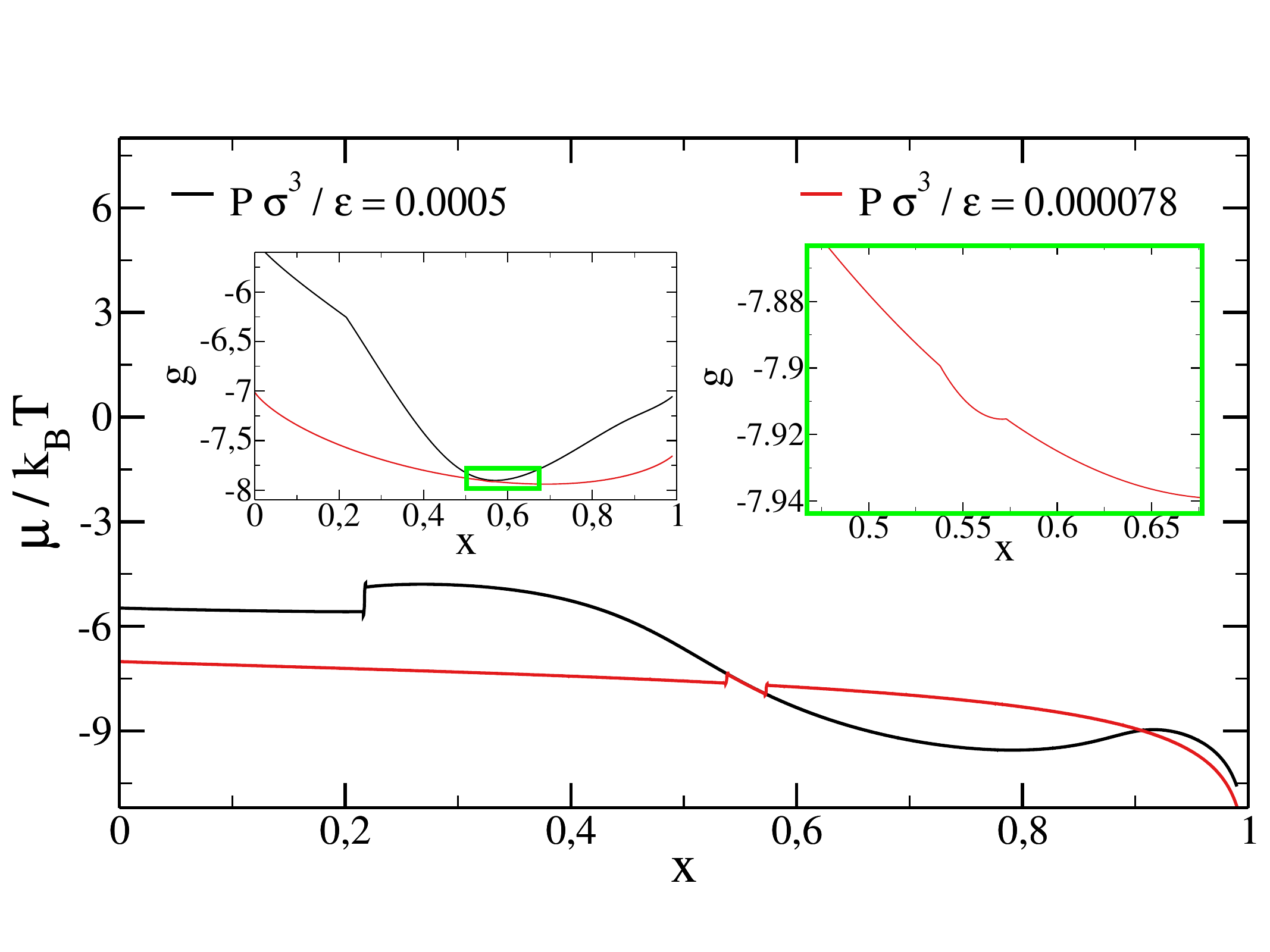}
    \caption{\textbf{Chemical potential behaviors near and far from azeotropic conditions.} The iterative process for shifting the azeotrope to a desired concentration follows a trial and error approach. In the graph we illustrate a step of the algorithm where setting  $\epsilon^{'}=1.2$ ultimately lead to an azeotrope at $x\sim0.55$. By changing the energy from $\epsilon$ to $\epsilon^{'}$ starting from a known azeotropic state  ($T_{azeo},P_{azeo},x_{azeo}$), the system moves away from the azeotropic conditions. The chemical potential as a function of the concentration, shown in black (or the Gibbs free energy curve (in black in the left inset) satisfying a common tangent construction) is tracked to identify when a new coexistence condition is met by modifying the temperature and the pressure. Next, by fine tuning the pressure, the thermodynamic conditions where the chemical potential becomes continuous and with a single concavity (corresponding to a tangency in the Gibbs free energy curves of the two phases) is progressively met, as illustrated by the red curves. The azeotropic concentration $x_{azeo}^{iter}$ at these new conditions ($T_{azeo}^{iter},P_{azeo}^{iter}$) is found and used as a feedback for the next iteration of $\epsilon^{'}$. 
    }
    \label{fig:algorithm}
\end{figure}

The azeotropic point can be systematically shifted through controlled modifications of interaction parameters. For each temperature and pressure, we theoretically construct the Gibbs free energy per particle $g(x)$ by minimizing Wertheim's free energy (Eq.~\ref{eqn:betaf_reference}) with respect to density for each composition $x$. Under coexistence conditions, $g(x)$ forms a two-branched curve representing the competing phases, with their intersection determining the equilibrium concentrations through the common tangent construction (see left inset of Fig.~\ref{fig:algorithm}). At the azeotropic point, these branches become tangent at a single concentration where both phases share identical composition.

From $g(x)$, we compute the chemical potentials $\mu_i(x)$, which exhibit distinct signatures of phase behavior. In coexistence regions, $\mu_i(x)$ becomes multivalued (with multiple concentrations yielding identical chemical potential) and develops discontinuities.
The width of these coexistence regions is pressure-dependent: it decreases with decreasing (increasing) pressure for mixtures exhibiting negative (positive) azeotropes \cite{smith1949introduction,Moore1962Physical}, and vanishes at the azeotropic point, as illustrated in Fig.~\ref{fig:algorithm}.

Our  algorithm begins with known azeotropic conditions $(P_{\text{azeo}}, T_{\text{azeo}}, x_{\text{azeo}})$. We apply small perturbations to the target bond energy $\epsilon$, then compute the updated $\mu_i(x)$. By adjusting pressure and temperature, we track the evolution of the coexistence region until $\mu_i(x)$ becomes single-valued, identifying the new azeotropic point $(P_{\text{azeo}}^{\text{iter}}, T_{\text{azeo}}^{\text{iter}}, x_{\text{azeo}}^{\text{iter}})$. This process iterates until convergence to the desired concentration $x_{\text{azeo}}^{\text{goal}}$ is achieved.

Applying this methodology to the N2c8 binary mixture, we predict that modifying the interaction parameter $\epsilon_{22} = \epsilon_{33}$ to $\epsilon' = 1.35$ should relocate the azeotropic point from $x = 0.5$ to $x = 0.6$. This prediction is verified through the following  theoretical and numerical analysis.

\begin{figure}[!t]
    \centering
    \includegraphics[width=0.5\textwidth]{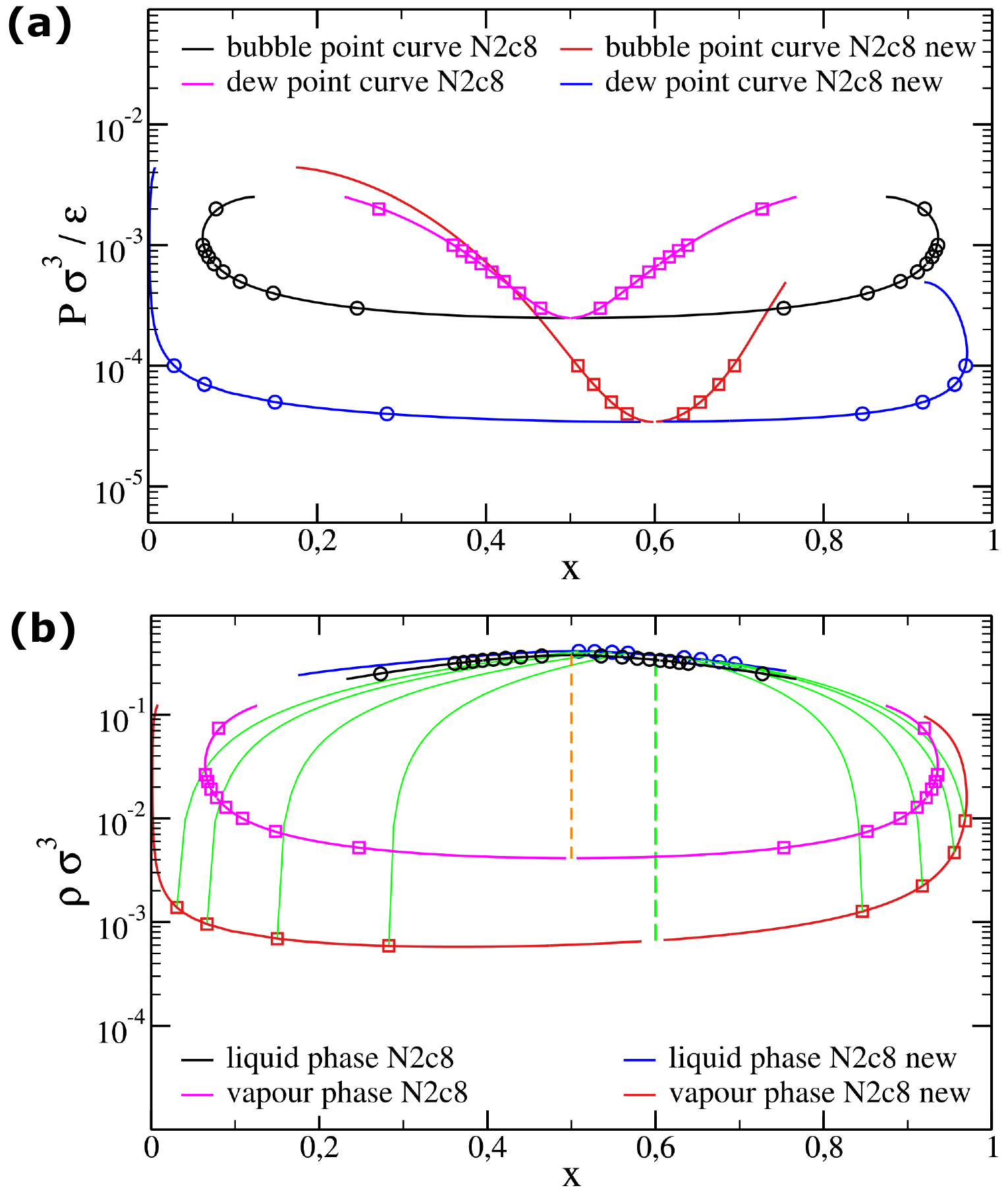}
    \caption{\textbf{Phase diagrams of N2c8 mixtures for different bond-energy.} %
    Phase diagrams comparing two interaction schemes at $T=0.08$: (1) uniform $\epsilon_i=1$ (black-magenta) and (2) differentiated $\epsilon_i=1$, $\epsilon_j=1.35$ for self-complementary pairs (blue-red). The azeotropic point shifts from $x=0.5$ to $x=0.6$ through selective bond energy modification while preserving topology. (a) Pressure-concentration diagram shows the azeotrope as the point where coexistence region collapses. (b) Density-concentration diagram identifies the azeotrope via vertical tie-lines (dashed).}
    \label{fig:wertheim_new}
\end{figure}

First, we compute the pressure-concentration and density-concentration phase diagrams within Wertheim's theory (Fig.~\ref{fig:wertheim_new}a and \ref{fig:wertheim_new}b). For comparison, these diagrams are superimposed with those of the original N2c8 mixture. The theoretical results demonstrate that while the original system shows azeotropic behavior at $x = 0.5$ (manifested as a single point where the coexistence region collapses), the modified system with $\epsilon' = 1.35$ exhibits this azeotropic point at $x = 0.6$. This shift is equally evident in the density-concentration diagram, where the characteristic straight tie-line (dashed) moves accordingly from $x = 0.5$ to $x = 0.6$.

\begin{figure}[!t]
    \centering
    \includegraphics[width=0.9\linewidth]{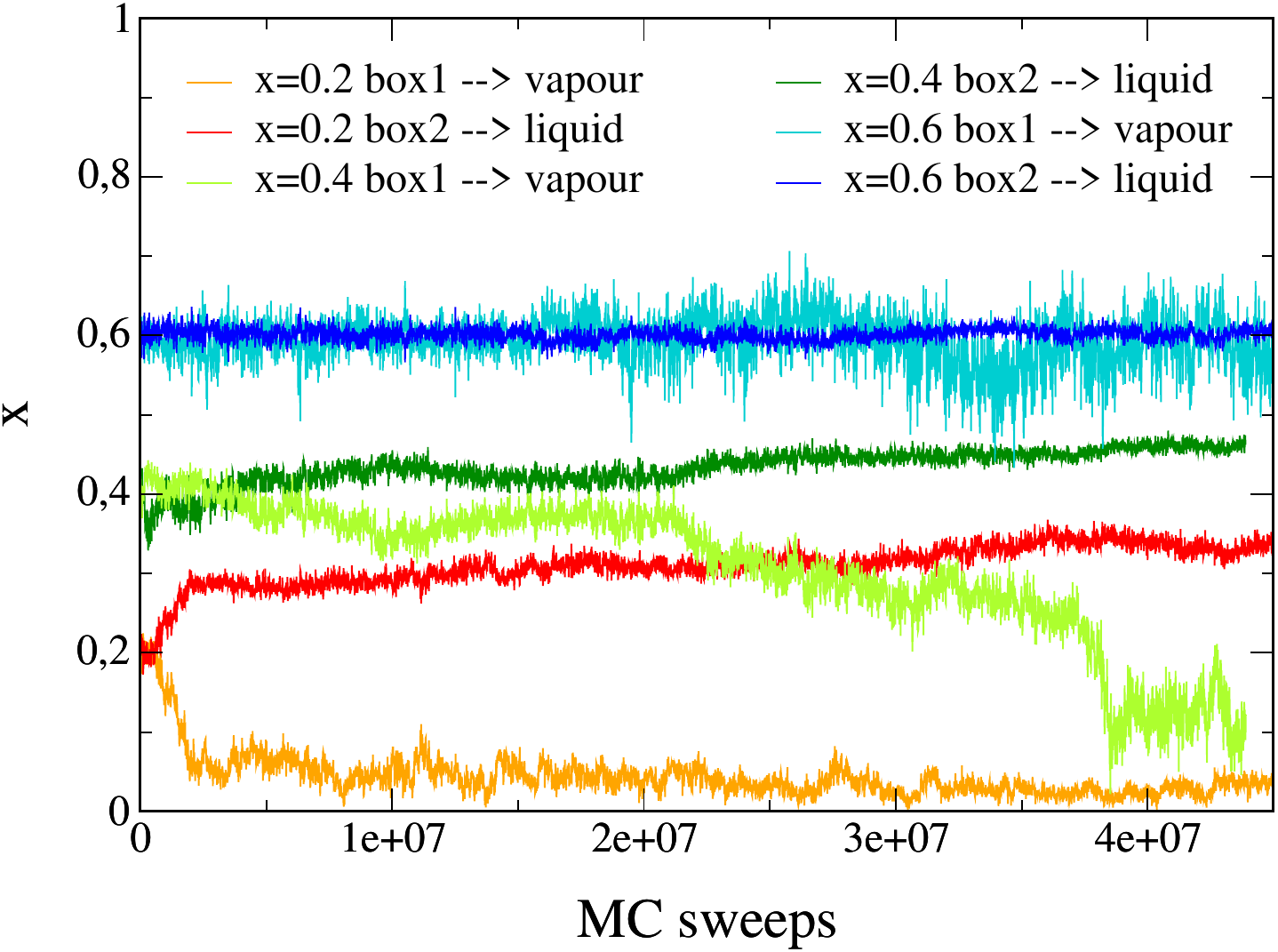}
    \caption{\textbf{Gibbs ensemble simulations for the new N2c8 binary mixture.} For different concentration of the first species ($x=0.2$, $x=0.4$, $x=0.6$), we initialized two equal boxes with density $\rho=0.1$, temperature $T=0.098$ and $N=500$ particles. By tracking the concentration of the first species throughout the simulations, we observe that the equilibrium value remains the same as the initial one when the two boxes are prepared at $x=0.6$.}
    \label{fig:gibbs_new}
\end{figure}

To validate these theoretical predictions, we perform Gibbs ensemble Monte Carlo simulations at temperature $T = 0.098$ and density $\rho = 0.1$. Systems are prepared with varying initial concentrations ($x = 0.2, 0.4, 0.6, 0.8$) in two equal simulation boxes. As shown in Fig.~\ref{fig:gibbs_new}, only the system with $x = 0.6$ maintains constant concentration in both boxes throughout the simulation, confirming this composition as the new azeotropic point. This agreement between theoretical predictions and numerical simulations provides robust verification of our design strategy.

\section{Conclusions}

In this work we implement an inverse thermodynamics approach to designing target thermodynamic behaviors through precise manipulation of microscopic interactions. We demonstrate this approach by forcing azeotropy in binary patchy particle mixtures, both via theoretical and numerical calculations.

Our focus on azeotropy is motivated by its fundamental role in self-assembly processes. By relocating the azeotropic point to match the stoichiometry of a target crystal, we optimize two-step nucleation pathways: the coexisting liquid phase naturally adopts the ideal concentration for templating the crystal, significantly boosting assembly rates~\cite{beneduce2023two}. We achieve this control through complementary approaches involving both bond topology design, which alters the connectivity rules between components, and bond energy modulation, which selectively tunes interaction strengths. Moreover, we identify an ideal azeotrope regime where demixing occurs azeotropically at all concentrations.

Methodologically, we combine Wertheim's thermodynamic perturbation theory with Gibbs ensemble simulations. This approach avoids computationally expensive full phase diagram calculations; instead, just a handful of targeted simulations suffice to validate azeotropic conditions by confirming concentration stability under coexistence.

Our work aims to bridge the gap between forward and inverse thermodynamics. By establishing design rules for azeotropicity, we provide a blueprint for tailoring phase behavior in self-assembling systems. More broadly, our approach exemplifies how Soft Matter’s tunability can be harnessed to solve thermodynamic challenges that we face when dealing with multi-component systems.

The principles of inverse thermodynamics, which enable precise control of phase behavior, can synergistically enhance solutions to the inverse self-assembly problem of designing target structures~\cite{whitelam2015statistical,bupathy2022temperature,bassani2024nanocrystal,hayakawa2024symmetry,michelson2025scalable}. Crucially, both inverse thermodynamics and inverse self-assembly can be formulated through Boolean satisfiability constraints on the building blocks' interactions. This unified computational framework, termed \emph{SAT-assembly}~\cite{romano2020designing,russo2022sat}, has already demonstrated remarkable success in engineering diverse nanostructures, including novel crystals, viral capsids, quasicrystals, and amorphous materials~\cite{rovigatti2022simple,pinto2023design,pinto2024automating}. Together, these advances establish a paradigm for \emph{materials by design}, where thermodynamic pathways and structural outcomes can be programmed in tandem to unlock new functional materials.

\section{Acknowledgments}
C.B., J.R., and F.S. acknowledge support by ICSC – Centro Nazionale di Ricerca in High Performance Computing, Big Data and Quantum Computing, funded by European Union – NextGenerationEU, and CINECA-ISCRA for HPC resources. Work by P.\v{S}. was supported by the U.S. Department of Energy (DOE), Office of Science, Basic Energy Sciences (BES) under Award No. DE-SC0025265.  We dedicate this manuscript to Athanassios Z. Panagiotopoulos
in honor of his 65th birthday.

\section{Author information}

\subsection{Notes}
The authors declare no competing financial interest.

\end{document}